\documentclass[aps,prd,superscriptaddress,showkeys,showpacs,a4paper,nofootinbib]{revtex4-2}

\usepackage[english]{babel}
\usepackage{amsmath,bm}
\usepackage{amssymb}
\usepackage{enumitem}
\usepackage{textcomp}
\usepackage{graphicx,float}
\usepackage{dsfont}
\usepackage{color}
\definecolor{G}{rgb}{0,0.6,0}

\usepackage[breaklinks=true]{hyperref}
\hypersetup{
	colorlinks=true,
	linkcolor=blue,
	filecolor=magenta,      
	urlcolor=blue,
	citecolor=red,
}

\usepackage{mathrsfs}
\usepackage[normalem]{ulem}

\usepackage{placeins}

\begin{document}
\large

\title{Hydrodynamical approach to chirality production during axion inflation}
\author{E.V.~Gorbar}
\affiliation{Physics Faculty, Taras Shevchenko National University of Kyiv, 64/13, Volodymyrska Street, 01601 Kyiv, Ukraine}
\affiliation{Bogolyubov Institute for Theoretical Physics, 14-b, Metrolohichna Street, 03143 Kyiv, Ukraine}

\author{A.I.~Momot}
\email{momot.andriy@knu.ua}
\affiliation{Physics Faculty, Taras Shevchenko National University of Kyiv, 64/13, Volodymyrska Street, 01601 Kyiv, Ukraine}

\author{O.O.~Prikhodko}
\affiliation{Physics Faculty, Taras Shevchenko National University of Kyiv, 64/13, Volodymyrska Street, 01601 Kyiv, Ukraine}

\author{O.M.~Teslyk}
\affiliation{Physics Faculty, Taras Shevchenko National University of Kyiv, 64/13, Volodymyrska Street, 01601 Kyiv, Ukraine}

\date{\today}

\begin{abstract}
	We study chirality production in the pseudoscalar inflation model of magnetogenesis taking into account the Schwinger effect and particle collisions in plasma in the relaxation time approximation. We consider the Schwinger production of one Dirac fermion species by an Abelian gauge field in two cases: (i)~the fermion carries only the weak charge with respect to the U(1) group and (ii)~it is also charged with respect to another strongly coupled gauge group. While the gradient-expansion formalism is employed for the description of the evolution of gauge field, plasma is described by hydrodynamical approach which allows us to determine the number, energy density, and chirality of produced fermions. It is found that while chirality production is very efficient for both, weakly and strongly interacting fermions, the resulting gauge field is typically stronger in the case of strongly interacting fermions due to suppression of the Schwinger conductivity by particle collisions.
\end{abstract}

\maketitle

\section{Introduction}
\label{sec:introduction}

The gamma-ray observations of blazars \cite{Neronov:2010,Taylor:2011,Tavecchio:2010,Caprini:2015} imply the lower bound on the strength of the present large-scale magnetic fields $B_{0}$ given by $B_{0}\gtrsim 10^{-16}\,\text{G}$ with coherence length possibly exceeding $1\,\text{Mpc}$. These observations strongly motivate the study of inflationary models of magnetogenesis because such models naturally explain the very large coherence length of generated magnetic fields.

Among various inflationary models of magnetogenesis the axion inflation model is especially attractive because it produces maximally helical magnetic fields. This is an advantageous characteristic because the survival of helical magnetic fields is more efficient in the primordial plasma compared to the case of nonhelical magnetic fields~\cite{Banerjee:2004}. The axion inflation model is characterized by the axial coupling of the inflaton field $\phi$ to the electromagnetic field by means of the interaction term $\beta\phi (\boldsymbol{E}\cdot\boldsymbol{B})/M_{\mathrm{P}}$, where $M_{\mathrm{P}}$ is the reduced Planck mass and $\beta$ is dimensionless coupling constant.

Obviously, in view of the chiral anomaly $\partial_{\mu}j^{\mu}_5=e^2/(2\pi^2)(\boldsymbol{E}\cdot\boldsymbol{B})$, axion inflation inevitably leads to chirality production. In turn, nonzero chiral density via the chiral magnetic effect~\cite{CME} affects the electromagnetic field evolution. Thus, axion inflation magnetogenesis implies a coupled evolutionary dynamics of the electromagnetic field and chirality.

A first step in the study of this joint evolution was done in Ref.~\cite{chirality-production}, where it was found that chirality production is indeed very efficient leading to the generation of a large chemical potential $\mu_5$ at the end of axion inflation. To analyze the evolution of the electromagnetic field, the gradient-expansion formalism~\cite{Sobol:2019} was employed accounting for the chiral magnetic effect~\cite{CME} via an additional term $\boldsymbol{j}_{\mathrm{CME}}=e^2/(2\pi^2)\mu_5\boldsymbol{B}$ in the expression for the electric current. Such a contribution to the electric current is induced in chirally asymmetric ultrarelativistic fermion plasma in a magnetic field.

Reference~\cite{chirality-production} used the simple expressions for the electric conductivity $\sigma$ induced by the Schwinger effect~\cite{Sauter:1931,Heisenberg:1936,Schwinger:1951} in collinear electric and magnetic fields in de Sitter spacetime~\cite{Domcke:2019} and assumed that a local thermodynamic equilibrium was reached. According to the discussion in Ref.~\cite{Domcke:2019}, such an assumption is questionable in the inflationary expanding Universe.

In the present paper, we extend the analysis performed in~\cite{chirality-production} in a few directions. First of all, we take into account the fact that the Schwinger induced current must be split into two contributions~\cite{Domcke:2018} where the first contribution originates from the lowest Landau level and, thus, corresponds to the chiral magnetic effect, while the second captures the contributions from higher Landau levels and is described by the usual Ohmic conductivity.

Secondly, we derive the Schwinger production rate $\Gamma$ and fermion energy production rate $\Gamma_{\rho}$ which enter the equations of motion for the fermion number and energy densities taking into account the fact that the electric and magnetic fields generated during axion inflation are not completely collinear. However, performing a Lorentz boost one can find an inertial frame where these fields are collinear. Then using the expressions for $\Gamma$ and $\Gamma_{\rho}$ for collinear fields~\cite{Dunne:2004,Ruffini:2009} and returning to the comoving reference frame, we obtain the corresponding expressions which account for noncollinearity of electric and magnetic fields generated during axion inflation.

The Schwinger effect~\cite{Sauter:1931,Heisenberg:1936,Schwinger:1951} of charged particle-antiparticle pairs production by a strong electric field is important for inflationary magnetogenesis because the generated electric fields are as strong as or even larger than the produced magnetic fields. We would like to note also that the energy production rate was not taken into account in \cite{chirality-production}. However, the analysis in this paper has shown that it plays an important role for the evolution of plasma and electromagnetic field. Finally, we have analyzed the impact of particle collisions on axion inflation magnetogenesis via the standard expression for the electric conductivity in the relaxation time approximation.

The paper is organized as follows. The axion inflation model is introduced in Sec.~\ref{sec:axion-infl}. The gradient-expansion formalism is described in Sec.~\ref{sec:axion-infl}. The set of equations for the evolution of plasma in the phenomenological hydrodynamical approach is derived in Sec.~\ref{sec:hydrodynamics}. The Schwinger pair and energy production rates are considered in Sec.~\ref{sec:pair-production}. The final set of equations is presented in Sec.~\ref{sec:final}. Numerical results for the chirality production and generated electromagnetic fields in the pseudoscalar inflation model are presented in Sec.~\ref{sec:analysis}. Section~\ref{sec:summary} is devoted to conclusions.

\section{Axion inflation model}
\label{sec:axion-infl}

In the axion inflation model the inflaton field is represented by a pseudoscalar (axionlike) field $\phi$ which is coupled to an Abelian gauge field $A_\mu$ through the term of the Chern-Simons type. The corresponding action reads as
\begin{equation}
	\label{S}
	S=\int d^{4}x\sqrt{-g} \left[\frac{1}{2}g^{\mu\nu}\,\partial_{\mu}\phi\,\partial_{\nu}\phi - V(\phi)-\frac{1}{4}F_{\mu\nu}F^{\mu\nu}-\frac{1}{4}I(\phi)F_{\mu\nu}\tilde{F}^{\mu\nu} +\mathcal{L}_{\mathrm{ch}}(\chi,\,A_{\mu})\right]\, ,
\end{equation}
where $g=\operatorname{det}g_{\mu\nu}$ is the determinant of the spacetime metric, $V(\phi)$ is the inflaton potential, $I(\phi)$ is the axial-coupling function, $F_{\mu\nu}=\partial_{\mu}A_{\nu}-\partial_{\nu}A_{\mu}$ is the gauge-field strength tensor, and
\begin{equation}
	\label{F}
	\tilde{F}^{\mu\nu}=\frac{1}{2\sqrt{-g}}\,\varepsilon^{\mu\nu\lambda\rho}F_{\lambda\rho}
\end{equation}
is the corresponding dual tensor; $\varepsilon^{\mu\nu\lambda\rho}$ is the absolutely antisymmetric Levi-Civita symbol with $\varepsilon^{0123}=+1$. The last term in Eq.~\eqref{S} is the Lagrangian for a generic matter field $\chi$ charged under the $U(1)$ gauge group and, therefore, coupled to the gauge field four-potential $A_{\mu}$. For the sake of generality, we will not specify this term and assume that it describes all charged fields in the model. 

Action~\eqref{S} implies the following Euler-Lagrange equations for the inflaton and gauge field:
\begin{equation}
	\label{L_E_1}
	\frac{1}{\sqrt{-g}}\partial_{\mu} \left[\sqrt{-g}\, g^{\mu\nu}\,\partial_{\nu}\phi \right] + \frac{dV}{d\phi} + \frac{1}{4}\frac{dI}{d\phi}F_{\mu\nu}\tilde{F}^{\mu\nu}=0\, ,
\end{equation}
\begin{equation}
	\label{L_E_2}
	\frac{1}{\sqrt{-g}}\partial_{\mu}\left[\sqrt{-g}\,F^{\mu\nu} \right]+ \frac{dI}{d\phi}\,\tilde{F}^{\mu\nu}\partial_{\mu}\phi=j^{\nu}\, ,
\end{equation}
where 
\begin{equation}
	j^{\nu}=-\frac{\partial \mathcal{L}_{\mathrm{ch}}(\chi,\,A_{\mu})}{\partial A_{\nu}}
\end{equation}
is the electric four-current. Equation \eqref{L_E_2} should be supplemented by the Bianchi identity for the dual gauge field strength tensor
\begin{equation}
	\label{Bianchi}
	\frac{1}{\sqrt{-g}}\partial_{\mu}\left[\sqrt{-g}\,\tilde{F}^{\mu\nu} \right] = 0\, .
\end{equation}

The energy-momentum tensor equals
\begin{equation}
	\label{T}
	T_{\mu\nu}=\frac{2}{\sqrt{-g}}\,\frac{\delta S}{\delta g^{\mu\nu}}=\partial_{\mu}\phi\,\partial_{\nu}\phi - g^{\lambda\rho}F_{\mu\lambda}F_{\nu\rho}-g_{\mu\nu}\left[\frac{1}{2}\partial_{\alpha}\phi\,\partial^{\alpha}\phi-V(\phi)-\frac{1}{4}F_{\alpha\beta}F^{\alpha\beta}\right]+T_{\mu\nu}^{\chi}\, ,
\end{equation}
where the last term describes the contribution of charged matter fields. 

The inflationary stage of the early Universe is characterized by the Friedmann--Lemaitre--Robertson--Walker metric $g_{\mu\nu}=(1,-a^2(t),-a^2(t),-a^2(t))$, where $a(t)$ is the scale factor. Further, it is convenient to use in the analysis of inflationary magnetogenesis the temporal gauge for the vector potential $A_{\mu}$, where $A_{\mu}=(0,\,-\boldsymbol{A})$. Then the three-vectors of electric $\bm{E}=(E^{1},\,E^{2},\,E^{3})$ and magnetic $\bm{B}=(B^{1},\,B^{2},\,B^{3})$ fields can be defined as $\bm{E}=-\tfrac{1}{a}\partial_{0}\bm{A}$ and $\bm{B}=\tfrac{1}{a^{2}}\operatorname{rot}\bm{A}$. They are physical fields measured by a comoving observer; therefore, we included the scale factor in their definition. The gauge field stress tensor and its dual tensor are expressed in terms of electric and magnetic fields as follows:
\begin{equation}
	F_{0i}=aE^{i}, \quad F_{ij}=-a^{2}\varepsilon_{ijk} B^{k}, \quad \tilde{F}_{0i}=aB^{i}, \quad \tilde{F}_{ij}=a^{2}\varepsilon_{ijk} E^{k}\, .
\end{equation}

The cosmic expansion rate (the Hubble parameter $H=\dot{a}/a$) is determined by the Friedmann equation
\begin{equation}
	\label{Friedmann}
	H^{2}=\frac{\rho}{3 M_{\mathrm{P}}^{2}}\, ,
\end{equation}
where the total energy density $\rho$ is given by the zero-zero component of the energy-momentum tensor \eqref{T},
\begin{equation}
	\label{rho}
	\rho=T_{0}^{0}=\left[\frac{1}{2}\dot{\phi}^{2}+V(\phi) \right] + \frac{1}{2}\left\langle \bm{E}^{2}+\bm{B}^{2} \right\rangle +\rho_{\mathrm{c}}\, .
\end{equation}
Here the two terms in square brackets correspond to the energy density of spatially homogeneous inflaton field, the next term describes the gauge field contribution (angular brackets denote the vacuum expectation value), and the last term is the counterpart for the charged matter fields. 

The electric four-current can be represented as
\begin{equation}
	j^{\mu}=\big(\rho_{\mathrm{ch}},\,\frac{1}{a}\bm{J}\big)\, .
\end{equation}
We assume that charged particles were absent in the Universe initially and were produced later in particle-antiparticle pairs via the Schwinger effect by a strong electric field connected with inflationary magnetogenesis. Therefore, we set the charge density to zero, $\rho_{\mathrm{ch}}=0$. On the other hand, the current density three-vector $\bm{J}$ may be nonzero in the presence of gauge field. Then the equations of motion \eqref{L_E_1}--\eqref{L_E_2} and \eqref{Bianchi} take the following form in the three-vector notation:
\begin{equation}
	\label{KGF}
	\ddot{\phi}+3H\dot{\phi}+V'(\phi)=I'(\phi)\left\langle \bm{E}\cdot\bm{B} \right\rangle\, ,
\end{equation}
\begin{equation}
	\label{Maxwell_1}
	\dot{\bm{E}}+2 H \bm{E}-\frac{1}{a} \operatorname{rot} \bm{B} + I'(\phi)\,\dot{\phi}\,\bm{B}+\bm{J}=0\, ,
\end{equation}
\begin{equation}
	\label{Maxwell_2}
	\dot{\bm{B}}+2 H \bm{B}+\frac{1}{a} \operatorname{rot} \bm{E}=0\,,
\end{equation}
\begin{equation}
	\label{Maxwell_3}
	\operatorname{div} \bm{E}=0\, , \qquad \operatorname{div} \bm{B}=0\, .
\end{equation}

Finally, to close the system of Maxwell`s equations, we need to specify the electric current. We will show below that it can be represented in the form of generalized Ohm's law,
\begin{equation}
	\label{Ohms-law}
	\bm{J}=\sigma_E\bm{E}+\sigma_B \bm{B}\, ,
\end{equation}
where $\sigma_E$ is the electric conductivity and $\sigma_B$ is related to the chiral magnetic effect. We will derive expressions for these conductivities below. The Maxwell equation \eqref{Maxwell_1} takes the form
\begin{equation}
	\label{Maxwell_1a}
	\dot{\bm{E}}+[2 H+\sigma_E] \bm{E} + [I'(\phi)\,\dot{\phi}+\sigma_B]\bm{B}-\frac{1}{a} \operatorname{rot} \bm{B}=0\, .
\end{equation}
As mentioned in the introduction, we employ the gradient-expansion formalism to solve the Maxwell equations in an inflating Universe.

\section{Gradient-expansion formalism}
\label{sec:GEF}

The gradient expansion formalism proposed in Ref.~\cite{Gorbar:2021} allows one to describe self-consistently the gauge field production in the axial inflation model including the Schwinger effect and the backreaction of produced particles on the gauge field. This method works in position space and, instead of vector quantities $\bm{E}$ and $\bm{B}$, operates with a set of scalar functions in a form of vacuum expectation values of scalar products of $\bm{E}$ and/or $\bm{B}$ with an arbitrary number of spatial derivatives (curls). These functions are
\begin{equation}
	\label{E_n}
	\mathscr{E}^{(n)}=\frac{1}{a^{n}}\left\langle \bm{E}\cdot \operatorname{rot}^{n} \bm{E}  \right\rangle\, ,
\end{equation}
\begin{equation}
	\label{G_n}
	\mathscr{G}^{(n)}=-\frac{1}{2a^{n}}\left\langle \bm{E}\cdot \operatorname{rot}^{n} \bm{B} + (\operatorname{rot}^{n} \bm{B}) \cdot \bm{E}  \right\rangle\, ,
\end{equation}
\begin{equation}
	\label{B_n}
	\mathscr{B}^{(n)}=\frac{1}{a^{n}}\left\langle \bm{B}\cdot \operatorname{rot}^{n} \bm{B}  \right\rangle\, .
\end{equation}

Using Eqs.~\eqref{Maxwell_2} and \eqref{Maxwell_1a}, we get the following equations of motion for bilinear electromagnetic functions:
\begin{equation}
	\label{dot_E_n}
	\dot{\mathscr{E}}^{(n)} + [(n+4)H+2\sigma_E]\,	\mathscr{E}^{(n)} - 2[I'(\phi)\dot{\phi}+\sigma_B]\,\mathscr{G}^{(n)} +2\mathscr{G}^{(n+1)}=[\dot{\mathscr{E}}^{(n)}]_{\mathrm{b}}\, ,
\end{equation}
\begin{equation}
	\label{dot_G_n}
	\dot{\mathscr{G}}^{(n)} +[(n+4)H+\sigma_E]\, \mathscr{G}^{(n)} - [I'(\phi)\dot{\phi}+\sigma_B]\,\mathscr{B}^{(n)}-\mathscr{E}^{(n+1)}+\mathscr{B}^{(n+1)}=[\dot{\mathscr{G}}^{(n)}]_{\mathrm{b}}\, ,
\end{equation}
\begin{equation}
	\label{dot_B_n}
	\dot{\mathscr{B}}^{(n)} + (n+4)H\,	\mathscr{B}^{(n)}-2\mathscr{G}^{(n+1)}=[\dot{\mathscr{B}}^{(n)}]_{\mathrm{b}}\, .
\end{equation}
The right-hand sides of Eqs.~\eqref{dot_E_n}--\eqref{dot_B_n} contain contributions due to boundary terms. Their necessity is dictated by the fact that the number of physically relevant modes (beyond the horizon) constantly grows during inflation. They have the following form:
\begin{equation}
	\label{E_p_d}
	[\dot{\mathscr{E}}^{(n)}]_{\mathrm{b}}=\frac{d \ln k_{\mathrm{h}}(t)}{d t}\frac{1}{4\pi^{2}}\left(\frac{k_{\mathrm{h}}(t)}{a(t)}\right)^{n+4}\sum_{\lambda=\pm 1}\lambda^{n} E_{\lambda}(\xi_{\mathrm{eff}}(t),s(t))\, ,
\end{equation}
\begin{equation}
	\label{G_p_d}
	[\dot{\mathscr{G}}^{(n)}]_{\mathrm{b}}=\frac{d \ln k_{\mathrm{h}}(t)}{d t}\frac{1}{4\pi^{2}}\left(\frac{k_{\mathrm{h}}(t)}{a(t)}\right)^{n+4}\sum_{\lambda=\pm 1}\lambda^{n+1}G_{\lambda}(\xi_{\mathrm{eff}}(t),s(t))\, ,
\end{equation}
\begin{equation}
	\label{B_p_d}
	[\dot{\mathscr{B}}^{(n)}]_{\mathrm{b}}=\frac{d \ln k_{\mathrm{h}}(t)}{d t}\frac{1}{4\pi^{2}}\left(\frac{k_{\mathrm{h}}(t)}{a(t)}\right)^{n+4}\sum_{\lambda=\pm 1}\lambda^{n}B_{\lambda}(\xi_{\mathrm{eff}}(t),s(t))\, ,
\end{equation}
where $k_{\mathrm{h}}(t)$ is the momentum of a mode which is crossing the horizon at the moment of time $t$:
\begin{equation}
	\label{k-h}
	k_{\mathrm{h}}(t)=\underset{t'\leq t}{\mathrm{max}}\Big\{a(t')H(t')\big[|\xi_{\mathrm{eff}}(t')|+\sqrt{\xi_{\mathrm{eff}}^{2}(t')+s^{2}(t')+s(t')}\big]\Big\}\,.
\end{equation}
Here we introduced the following parameters:
\begin{equation}
	\label{xi-s}
	\xi_{\mathrm{eff}}(t)=\frac{dI}{d\phi}\frac{\dot{\phi}}{2H}+\frac{\sigma_B(t)}{2H}\, , \qquad s(t)=\frac{\sigma_E(t)}{2H}\, .
\end{equation}
The functions $E_{\lambda}$, $G_{\lambda}$, and $B_{\lambda}$ were derived in Ref.~\cite{Gorbar:2021} and have the form
\begin{multline}
	\label{E-lambda}
	E_{\lambda}(\xi_{\mathrm{eff}},s)=\frac{e^{\pi\lambda \xi_{\mathrm{eff}}}}{r^{2}(\xi_{\mathrm{eff}},s)} \Big|\left(i r(\xi_{\mathrm{eff}},s)-i\lambda \xi_{\mathrm{eff}}-s\right)W_{-i\lambda\xi_{\mathrm{eff}},\frac{1}{2}+s}(-2i r(\xi_{\mathrm{eff}},s))+\\+W_{1-i\lambda\xi_{\mathrm{eff}},\frac{1}{2}+s}(-2i r(\xi_{\mathrm{eff}},s))\Big|^{2}\, ,
\end{multline}
\begin{multline}
	\label{G-lambda}
	G_{\lambda}(\xi_{\mathrm{eff}},s)=
	\frac{e^{\pi\lambda \xi_{\mathrm{eff}}}}{r(\xi_{\mathrm{eff}},s)} \bigg\{\Re e\left[W_{i\lambda \xi_{\mathrm{eff}},\frac{1}{2}+s}(2i r(\xi_{\mathrm{eff}},s)) W_{1-i\lambda\xi_{\mathrm{eff}},\frac{1}{2}+s}(-2i r(\xi_{\mathrm{eff}},s))\right]-\\-s\left|W_{-i\lambda\xi_{\mathrm{eff}},\frac{1}{2}+s}(-2i r(\xi_{\mathrm{eff}},s)) \right|^{2}\bigg\}\, ,
\end{multline}
\begin{equation}
	\label{B-lambda}
	\qquad B_{\lambda}(\xi_{\mathrm{eff}},s)=e^{\pi\lambda \xi_{\mathrm{eff}}}\,\left|W_{-i\lambda\xi_{\mathrm{eff}},\frac{1}{2}+s}(-2i r(\xi_{\mathrm{eff}},s)) \right|^{2}
\end{equation}
with $r(\xi_{\mathrm{eff}},s)=|\xi_{\mathrm{eff}}|+\sqrt{\xi_{\mathrm{eff}}^{2}+s+s^{2}}$.

Note that the equation of motion for the $n$th order function always contains at least one function with the $(n+1)$th power of the curl. As a result, all equations are coupled into an infinite chain that needs to be truncated in practice. The easiest way to perform such a truncation is to express higher order quantities in terms of the lower order ones. For some maximal order $n_{\mathrm{max}}$, we impose the following conditions:
\begin{equation}
	\label{truncation}
	X^{(n_{\mathrm{max}}+1)}\approx \Big(\frac{k_{\mathrm{h}}}{a}\Big)^{2} X^{(n_{\mathrm{max}}-1)}
\end{equation}
for $X=\{\mathscr{E},\,\mathscr{G},\,\mathscr{B}\}$. This truncation rule respects transformation properties of $X^{(n)}$ under parity (i.e., relates either scalars or pseudoscalars). The truncation order $n_{\mathrm{max}}$ must be chosen in such a way that its further increase does not lead to a significant change of the result.

\section{Hydrodynamical description of plasma}
\label{sec:hydrodynamics}

To describe the dynamics of particles produced due to the Schwinger effect and their backreaction on the gauge field, we use the hydrodynamical approach and define the corresponding system of hydrodynamical equations in this section.
Electromagnetohydrodynamics (EMHD)~\cite{Bisnovatyi-Kogan} is an extension of the familiar
magnetohydrodynamics (MHD)~\cite{Alfven} to the case where the displacement current and the generation of electromagnetic waves is important. Since the dynamics of gauge field is definitely very important in the study of inflationary magnetogenesis, we adopt in this section the EMHD approach for the description of produced particles.

An important assumption of EMHD (as well as more familiar MHD) is that the corresponding plasma is strongly collisional with the time scale of collisions shorter than the other characteristic times in the system. 
Certainly, the kinetic theory with its main ingredient in the form of the particle distribution function $f(\bm{p},\bm{x},t)$ in the phase space would provide more accurate approach to the description of plasma; however, it is not easy to solve the corresponding Boltzmann equation. Therefore, although the hydrodynamical approach is more coarse and may sometimes miss the important physics, it is relatively simple because the hydrodynamical variables depend only on spacetime coordinates $\bm{x}$ and $t$. In addition, the hydrodynamical approach captures many of the important properties of plasma dynamics and is often qualitatively correct. Therefore, we use in the present paper the EMHD approach paying a special attention to the problem of attaining the collisional regime in our numerical analysis and leaving the implementation of the kinetic approach for future studies.

Since we consider a spatially uniform system with vanishing pressure and temperature gradients, we could set the hydrodynamical velocity $\bm{v}(\bm{x},t)$ to zero. Then we are left with the following set of hydrodynamical variables which describe the plasma of chiral fermions (with negligibly small mass). These are the total number density of particles and antiparticles $n$, chirality density $n_5$, total energy density $\rho_{\mathrm{c}}$, and conduction current $\boldsymbol{j}_{\mathrm{cond}}$. Let us define now the set of equations which govern these hydrodynamical variables.

We begin with the equations of motion for the fermion number and energy densities. The Schwinger effect is characterized by the pair creation rate per unit volume and unit time $\Gamma$ and the energy production rate per unit volume and unit time $\Gamma_\rho$ which will be specified in the next section. The equations of motion for the fermion number and energy densities have the following form:
\begin{equation}
	\frac{dn}{dt}+3H n= 2\Gamma\, ,
\end{equation}
\begin{equation}
	\label{eq-rho}
	\frac{d\rho_{\mathrm{c}}}{dt}+4H \rho_{\mathrm{c}}= \Gamma_\rho+(\bm{E}\cdot\bm{j}_{\mathrm{cond}})\, ,
\end{equation}
where terms with $H$ take into account the redshift due to the Universe expansion. The first term on the right-hand side of Eq.~(\ref{eq-rho}) corresponds to the energy transfer from the gauge field to fermions due to the pair creation process while the second term describes the increase of energy of produced particles in an external electric field. The former term can also be described as the scalar product of the electric field with some effective current---the so called polarization current:
\begin{equation}
	\Gamma_\rho=(\bm{E}\cdot\bm{j}_{\mathrm{pol}})\, ,
\end{equation}
therefore, the equation of motion for the energy density can be rewritten as
\begin{equation}
	\label{eq-rho-2}
	\frac{d\rho_{\mathrm{c}}}{dt}+4H \rho_{\mathrm{c}}= (\bm{E}\cdot\bm{J})\, ,
\end{equation}
where 
\begin{equation}
	\bm{J}=\bm{j}_{\mathrm{pol}}+\bm{j}_{\mathrm{cond}}
\end{equation}
is the total electric current. Since we would like to combine the hydrodynamical approach with the gradient-expansion formalism, we represent the electric current in the form of the generalized Ohm's law:
\begin{equation}
	\bm{J}=\sigma_{E,\mathrm{total}} \bm{E}+\sigma_B \bm{B}\, .
\end{equation}
Then
\begin{equation}
	\sigma_{E,\mathrm{total}}=\sigma_{E}+\frac{\Gamma_\rho}{\langle\bm{E}^2\rangle}\, ,
\end{equation}
where the second term can be considered as the polarization conductivity. We will specify the explicit expressions for conductivities $\sigma_{E,B}$ below.

In order to write down the equation of motion for the chirality density, we note that the Schwinger pair production is insensitive to the chirality, \textit{i.e.}, it cannot produce the net chirality. However, since we have, in general, non-orthogonal electric and magnetic fields, the chiral anomaly leads to chirality production. Finally, the chirality flipping processes with the chirality flipping rate $\Gamma_{\mathrm{flip}}$ could lead to the equilibration of the chiral imbalance. Taking into account the above-mentioned effects, we find the following equation for the chirality density $n_5$:
\begin{equation}
	\label{eq-n5}
	\frac{dn_5}{dt}+3H n_5= \frac{e^2}{2\pi^2}\langle \boldsymbol{E}\cdot\boldsymbol{B}\rangle-\Gamma_{\mathrm{flip}} n_5\, .
\end{equation}
In the Standard Model, the chirality flipping rate $\Gamma_{\mathrm{flip}}$ is much less than the Hubble rate at temperatures above 80\,TeV~\cite{Campbell:1992jd,Bodeker:2019ajh}, therefore, we will neglect these processes in what follows.

\section{Schwinger pair and energy production rates}
\label{sec:pair-production}

In the previous section, we specified the equations which govern the temporal evolution of hydrodynamical variables. These equations depend on the Schwinger pair and energy production rates which we determine in this section. The Schwinger pair creation rate
\begin{equation}
	\Gamma=\frac{1}{V}\frac{dN_{\mathrm{pairs}}}{dt}
\end{equation}
is a Lorentz scalar and the energy production rate
\begin{equation}
	\Gamma_\rho=\frac{1}{V}\frac{dW}{dt}
\end{equation}
transforms as the 0th component of a contravariant four-vector.

As mentioned in the Introduction, although electric and magnetic fields generated during axion inflation are not completely collinear, performing a Lorentz boost one can find an inertial frame where these fields are collinear. Then, following the analysis in~\cite{Dunne:2004,Ruffini:2009}, one can easily calculate the Schwinger number and energy production rates in that frame and return back to the initial frame performing the corresponding Lorentz boost. This procedure allows us to take into account noncollinearity of electric and magnetic fields generated during axion inflation. Obviously, such a procedure implicitly assumes that the spatial dependence of electric and magnetic fields can be neglected.

The calculation of the Schwinger number and energy production rates in the reference frame where electric and magnetic fields are collinear is straightforward (we denote electric and magnetic fields in this reference frame with tilde). Then the production rates can be computed using the semiclassical approximation. For the Dirac fermion with mass $m$, the Schwinger pair production process can be regarded as quantum tunneling through the energy gap between the upper and lower continua. In the presence of constant and collinear electric and magnetic fields directed along the $z$-axis the semiclassical energy of the fermion has the form
\begin{equation}
	\mathcal{E}_{\pm}=|e\tilde{E}|z\pm \sqrt{p_z^2+2|e\tilde{B}|(n+\tfrac{1}{2}+\hat{\sigma})+m^2}\, ,
\end{equation}
where $n=0,\,1,\,2,\,\dots$ is the Landau level number, $\hat{\sigma}=\pm\tfrac{1}{2}$ is the spin projection on the $z$ axis, and $p_z$ is the continuous momentum along the $z$-axis. A fermion with energy $\mathcal{E}$ and given values of $n$ and $\hat{\sigma}$ can propagate in the regions of space where $p_z$ is real. However, there is a finite region where $p_z$ can be only imaginary which is the classically forbidden region. The tunneling probability is proportional to~\cite{Dunne:2004,Ruffini:2009}
\begin{equation}
	\mathcal{P}\propto \exp\bigg(-2\int\limits_{z_{-}}^{z_{+}}|p_z| dz \bigg)=\exp\bigg( -\pi\frac{2|e\tilde{B}|(n+\tfrac{1}{2}+\hat{\sigma})+m^2}{|e\tilde{E}|}\bigg)\, ,
\end{equation}
where $z_\pm$ are classical turning points. This expression gives the probability of the pair production at the $n$th Landau level with the spin projection $\hat{\sigma}$. Following Ref.~\cite{Ruffini:2009}, we can use it to compute the pair and energy production rates. For simplicity, let us consider the fermion with vanishing mass.
\footnote{The mass can be neglected if $m^2\ll |e\tilde{E}|$, which is typically satisfied during inflation. In the opposite case, the Schwinger pair production is exponentially suppressed and, therefore, is not interesting for the present study.}
Then we get
\begin{multline}
	\Gamma=\frac{|e\tilde{E}||e\tilde{B}|}{4\pi^2} \sum_{n,\hat{\sigma}} \exp\bigg( -\pi\frac{2|e\tilde{B}|(n+\tfrac{1}{2}+\hat{\sigma})}{|e\tilde{E}|}\bigg)=\\= \frac{|e\tilde{E}||e\tilde{B}|}{4\pi^2} \bigg[1+2\sum_{k=1}^{\infty}e^{-2\pi \frac{|\tilde{B}|}{|\tilde{E}|}k}\bigg]= \frac{|e\tilde{E}||e\tilde{B}|}{4\pi^2} \operatorname{coth}\Big(\pi\frac{|\tilde{B}|}{|\tilde{E}|}\Big)\, ,
\end{multline}
\begin{multline}
	\tilde{\Gamma}_\rho=\frac{|e\tilde{E}||e\tilde{B}|}{4\pi^2} \sum_{n,\hat{\sigma}} 2\sqrt{2|e\tilde{B}|(n+\tfrac{1}{2}+\hat{\sigma})}\exp\bigg( -\pi\frac{2|e\tilde{B}|(n+\tfrac{1}{2}+\hat{\sigma})}{|e\tilde{E}|}\bigg)=\\= \frac{|e\tilde{E}||e\tilde{B}|}{4\pi^2}4\sqrt{2|e\tilde{B}|}\sum_{k=1}^{\infty}\sqrt{k}e^{-2\pi \frac{|\tilde{B}|}{|\tilde{E}|}k}= \frac{|e\tilde{E}||e\tilde{B}|^{3/2}\sqrt{2}}{\pi^2} \operatorname{Li}_{-\frac{1}{2}}\Big(e^{-2\pi\frac{|\tilde{B}|}{|\tilde{E}|}}\Big)\, ,
\end{multline}
where $\operatorname{Li}_{-\frac{1}{2}}$ is the polylogarithm of order $-\frac{1}{2}$. In the second expression, factor $2\sqrt{2|e\tilde{B}|(n+\tfrac{1}{2}+\hat{\sigma})}$ is the energy difference between the positive and negative energy continua at fixed $z$.

As stated above, electric $\bm{E}$ and magnetic $\bm{B}$ fields generated during axion inflation are, in general, not collinear. Let us assume without loss of generality that $\bm{E}$ and $\bm{B}$ in the comoving frame lie in $xOy$ plane. Then the velocity of boost leading to the collinear frame is parallel to $z$ axis and equals to
\begin{equation}
	\boldsymbol{v}=\frac{2[\bm{E}\times \bm{B}]}{\bm{E}^2+\bm{B}^2+\sqrt{(\bm{E}^2-\bm{B}^2)^2+4(\bm{E}\cdot\bm{B})^2}}\, .
\end{equation}
The corresponding Lorentz factor is given by
\begin{equation}
	\gamma=\frac{1}{\sqrt{1-v^2}}=\frac{1}{\sqrt{2}}\bigg[1+\frac{\bm{E}^2+\bm{B}^2}{\sqrt{(\bm{E}^2-\bm{B}^2)^2+4(\bm{E}\cdot\bm{B})^2}}\bigg]^{1/2}.
\end{equation}

The resulting values of electric and magnetic fields in the collinear frame can be easily found from the invariants of the gauge-field tensor which are the same in both frames
\begin{equation}
	\mathcal{I}_1=\frac{1}{2} F_{\mu\nu}F^{\mu\nu} = \bm{B}^2 - \bm{E}^2=\tilde{B}^2-\tilde{E}^2\, ,
\end{equation}
\begin{equation}
	\mathcal{I}_2=-\frac{1}{4} F_{\mu\nu}\tilde{F}^{\mu\nu} = \bm{E}\cdot\bm{B} =\tilde{E}\,\tilde{B}
\end{equation}
that gives
\begin{equation}
	\tilde{E}=\frac{1}{\sqrt{2}}\bigg[\bm{E}^2-\bm{B}^2+\sqrt{(\bm{E}^2-\bm{B}^2)^2+4(\bm{E}\cdot\bm{B})^2}\bigg]^{1/2}\, ,
\end{equation}
\begin{equation}
	\tilde{B}=\mathrm{sign\,}(\bm{E}\cdot\bm{B})\frac{1}{\sqrt{2}}\bigg[\bm{B}^2-\bm{E}^2+\sqrt{(\bm{E}^2-\bm{B}^2)^2+4(\bm{E}\cdot\bm{B})^2}\bigg]^{1/2}\, .
\end{equation}
Note that $\tilde{E}$ is defined as a positive quantity while the projection of magnetic field on the direction of electric field, $\tilde{B}$, may have any sign depending on the scalar product $\bm{E}\cdot\bm{B}$ in the comoving frame.

Finally, expressing everything in terms of gauge fields in the comoving frame, we obtain the sought expression for the pair production rate per unit volume and unit time
\begin{equation}
	\Gamma=\frac{e^2|\bm{E}\cdot\bm{B}|}{4\pi^2}\operatorname{coth}\frac{2\pi |\bm{E}\cdot\bm{B}|}{\bm{E}^2-\bm{B}^2+\Delta},\quad \Delta=\sqrt{(\bm{E}^2-\bm{B}^2)^2+4(\bm{E}\cdot\bm{B})^2}\, .
\end{equation}

The energy production rate in the comoving frame can be found by performing the inverse Lorentz boost and is given by
\begin{multline}
	\Gamma_\rho=\gamma \tilde{\Gamma}_\rho=\frac{e^{5/2}|\bm{E}\cdot\bm{B}|}{2^{1/4}\pi^2}\frac{(\bm{E}^2+\bm{B}^2+\Delta)^{1/2}(\bm{B}^2-\bm{E}^2+\Delta)^{1/4}}{\Delta^{1/2}}
	\operatorname{Li}_{-\frac{1}{2}}\Big( e^{-\frac{4\pi |\bm{E}\cdot\bm{B}|}{\bm{E}^2-\bm{B}^2+\Delta}}\Big)=\\=
	\frac{e^{5/2}2^{1/4}}{\pi^2}\frac{(\bm{E}^2+\bm{B}^2+\Delta)^{1/2}}{\Delta^{1/2}(\bm{E}^2-\bm{B}^2+\Delta)^{1/4}}
	|\bm{E}\cdot\bm{B}|^{3/2}\operatorname{Li}_{-\frac{1}{2}}\Big( e^{-\frac{4\pi |\bm{E}\cdot\bm{B}|}{\bm{E}^2-\bm{B}^2+\Delta}}\Big)\, .
\end{multline}
Before proceeding to numerical analysis, it is useful to collect everything and present the final set of equations.

\section{Final set of equations}
\label{sec:final}

The system of equations governing the joint evolution of the inflaton field, gauge field, and produced plasma has the following form:
\begin{itemize}
	\item the Friedmann equation for the Hubble rate
	\begin{equation}
		\label{Friedmann-fin-1}
		H^2=\Big(\frac{\dot{a}}{a}\Big)^2=\frac{1}{3M_{\mathrm{P}}^2}\Big[\frac{1}{2}\dot{\phi}^2+V(\phi)+\frac{1}{2}\big(\mathscr{E}^{(0)}+\mathscr{B}^{(0)}\big)+\rho_{\mathrm{c}}\Big]\, ,
	\end{equation}
	
	\item the Klein--Gordon equation for the inflaton field
	\begin{equation}
		\label{KGF-fin-1}
		\ddot{\phi}+3H\dot{\phi}+V'(\phi)=-I'(\phi) \mathscr{G}^{(0)}\, ,
	\end{equation}
	
	\item the gradient-expansion formalism equations for the gauge-field bilinear functions
	\begin{multline}
		\label{dot_E_n-fin-1}
		\dot{\mathscr{E}}^{(n)} + \big[(n+4)H+2\sigma_{E}+2\frac{\Gamma_\rho}{\mathscr{E}^{(0)}}\big]\,	\mathscr{E}^{(n)} - 2[I'(\phi)\dot{\phi}+\sigma_B]\,\mathscr{G}^{(n)} +2\mathscr{G}^{(n+1)}=\\=\frac{d \ln k_{\mathrm{h}}(t)}{d t}\frac{1}{4\pi^{2}}\left(\frac{k_{\mathrm{h}}(t)}{a(t)}\right)^{n+4}\sum_{\lambda=\pm 1}\lambda^{n} E_{\lambda}(\xi_{\mathrm{eff}}(t),s(t))\, ,
	\end{multline}
	\begin{multline}
		\label{dot_G_n-fin-1}
		\dot{\mathscr{G}}^{(n)} +\big[(n+4)H+\sigma_{E}+\frac{\Gamma_\rho}{\mathscr{E}^{(0)}}\big]\, \mathscr{G}^{(n)} - [I'(\phi)\dot{\phi}+\sigma_B]\,\mathscr{B}^{(n)}-\mathscr{E}^{(n+1)}+\mathscr{B}^{(n+1)}=\\
		=\frac{d \ln k_{\mathrm{h}}(t)}{d t}\frac{1}{4\pi^{2}}\left(\frac{k_{\mathrm{h}}(t)}{a(t)}\right)^{n+4}\sum_{\lambda=\pm 1}\lambda^{n+1}G_{\lambda}(\xi_{\mathrm{eff}}(t),s(t))\, ,
	\end{multline}
	\begin{equation}
		\label{dot_B_n-fin-1}
		\dot{\mathscr{B}}^{(n)} + (n+4)H\,	\mathscr{B}^{(n)}-2\mathscr{G}^{(n+1)}=\frac{d \ln k_{\mathrm{h}}(t)}{d t}\frac{1}{4\pi^{2}}\left(\frac{k_{\mathrm{h}}(t)}{a(t)}\right)^{n+4}\sum_{\lambda=\pm 1}\lambda^{n}B_{\lambda}(\xi_{\mathrm{eff}}(t),s(t))\, ,
	\end{equation}
	where
	\begin{equation}
		\label{xi-s-fin-1}
		\xi_{\mathrm{eff}}(t)=\frac{dI}{d\phi}\frac{\dot{\phi}}{2H}+\frac{\sigma_B}{2H}\, , \qquad s(t)=\frac{\sigma_{E}+\Gamma_\rho/\mathscr{E}^{(0)}}{2H}\, ,
	\end{equation}
	
	\item the equation for the fermion energy density
	\begin{equation}
		\label{dot-rho_c-fin-1}
		\frac{d\rho_{\mathrm{c}}}{dt}+4H \rho_{\mathrm{c}}=\Gamma_\rho +\sigma_{E}\mathscr{E}^{(0)}-\sigma_{B}\mathscr{G}^{(0)}\, ,
	\end{equation}
	
	\item the equation for the fermion chirality density
	\begin{equation}
		\label{dot-n5-fin-1}
		\frac{dn_5}{dt}+3H n_5=- \frac{e^2}{2\pi^2}\mathscr{G}^{(0)}\, ,
	\end{equation}
	
	\item the equation for the fermion number density
	\begin{equation}
		\frac{dn}{dt}+3H n= 2\Gamma\, .
	\end{equation}
\end{itemize}

Here 
\begin{equation}
	\Gamma=\frac{e^2|\mathscr{G}^{(0)}|}{4\pi^2}\operatorname{coth}\frac{2\pi |\mathscr{G}^{(0)}|}{\mathscr{E}^{(0)}-\mathscr{B}^{(0)}+\Delta}\, ,
\end{equation}
\begin{equation}
	\Gamma_\rho=\frac{e^{5/2}2^{1/4}}{\pi^2}\frac{(\mathscr{E}^{(0)}+\mathscr{B}^{(0)}+\Delta)^{1/2}}{\Delta^{1/2}(\mathscr{E}^{(0)}-\mathscr{B}^{(0)}+\Delta)^{1/4}}
	|\mathscr{G}^{(0)}|^{3/2}\operatorname{Li}_{-\frac{1}{2}}\Big( e^{-\frac{4\pi |\mathscr{G}^{(0)}|}{\mathscr{E}^{(0)}-\mathscr{B}^{(0)}+\Delta}}\Big)
\end{equation}
with
\begin{equation}
	\Delta=\sqrt{(\mathscr{E}^{(0)}-\mathscr{B}^{(0)})^2+4(\mathscr{G}^{(0)})^2}\, .
\end{equation}

Finally, for conductivities, we take the expressions from Ref.~\cite{Domcke:2018}, see Eqs.~(4.11)--(4.12):
\begin{equation}
	\label{sigmaE_D}
	\sigma_E=\frac{e^3}{3\pi^2 H}\frac{\sqrt{\mathscr{B}^{(0)}}}{\exp\Big(2\pi\sqrt{\frac{\mathscr{B}^{(0)}}{\mathscr{E}^{(0)}}}\Big)-1}\, ,
\end{equation}
\begin{equation}
	\label{sigmaB_D}
	\sigma_B=-\frac{e^3}{6\pi^2 H}\sqrt{\mathscr{E}^{(0)}}\operatorname{sign}(\mathscr{G}^{(0)})\, .
\end{equation}

As we discussed in Sec.~\ref{sec:hydrodynamics}, the hydrodynamical approach implies that plasma is in the collisional regime with the time scale of collisions shorter than the other characteristic times in the system. The characteristic time scale of collisions $\tau$ of charged fermions in plasma at thermal equilibrium with temperature $T$ can be estimated as \cite{Thoma:2009a,Thoma:2009b}:
\begin{equation}
	\label{tau}
	\tau_{\mathrm{eq}}=\frac{c_0}{T\big(\frac{e^2}{4\pi}\big)^2 \ln|e|^{-1}}\, ,
\end{equation}
where $e=\sqrt{4\pi\alpha_{\mathrm{w}}}\approx 0.35$ is the gauge charge
\footnote{Here we used the value of the Standard Model hypercharge coupling constant $g'$ at the energy scale of the $Z$-boson mass, $m_Z=91.2\,$GeV. If view of the coupling constant running with momentum, better choice would be to take the value of the hypercharge coupling constant at the Hubble scale. However, the latter is model dependent, hence, the collision time is a model dependent quantity too.}
and $c_0$ is a dimensionless constant of order unity.
\footnote{In general, it depends on particle's momentum; however, for simplicity we will neglect this dependence and assume it to be constant. Its numerical value also depends on the number of charged degrees of freedom in plasma. We will keep it as a free parameter which determines the intensity of collisions in plasma.} 

Equation~\eqref{tau} can be used for the system in the state of thermodynamic equilibrium, in which the temperature can be introduced. In the beginning of the inflation the system is definitely not in the equilibrium state and it is not obvious that it will come to the equilibrium at the end of inflation. On the other hand, collisions still may play an important role during inflation. To deal with such a case, instead of Eq.~\eqref{tau}, we can estimate the collision time for charged particles as follows:
\begin{equation}
	\tau_{\mathrm{p}}=\frac{\rho_{\mathrm{c}}^2}{ke^4 \ln(e^{-1}) n^3}\, ,
\label{collision-time-leptons}
\end{equation}
where $k$ is a model dependent factor which accounts for the number of charged particle species and their interaction strength.
In our analysis, we consider the two limiting cases $k=1$ (for example, it could be a single lepton interacting only via electroweak interactions in the Standard Model) and $k=10^{4}$. The latter can be considered as the case of strongly interacting particles, e.g., quarks in the Standard Model, where $k$ equals the product of the number of quark species $N_{\mathrm{q}}$ and the square of the ratio of the strong and weak coupling constants $(\alpha_{\mathrm{s}}/\alpha_\mathrm{w})^2$. Clearly, factor $k$ is quite model dependent because neither the number of particle species nor their interaction strength are fixed at the Hubble scale of inflation. Therefore, our choice of two numerical values of $k$ is by no means definite but serves only for illustrative purposes.

If the collision time is less than Hubble time $\tau_H \simeq 1/H$, then the expressions for conductivities can be modified by replacing $1/(3H)$ with the collision time $\tau_{\mathrm{p}}$~\cite{Domcke:2018}
\begin{equation}
	\label{sigmaE}
	\sigma_E=\frac{e^3 \tau_{\mathrm{p}}}{\pi^2}\frac{\sqrt{\mathscr{B}^{(0)}}}{\exp\Big(2\pi\sqrt{\frac{\mathscr{B}^{(0)}}{\mathscr{E}^{(0)}}}\Big)-1}\, ,
\end{equation}
\begin{equation}
	\label{sigmaB}
	\sigma_B=-\frac{e^3 \tau_{\mathrm{p}}}{2\pi^2}\sqrt{\mathscr{E}^{(0)}}\operatorname{sign}(\mathscr{G}^{(0)})\, .
\end{equation}
Having presented the complete system of equations, we proceed now to its numerical analysis.

\section{Numerical analysis}
\label{sec:analysis}

We consider two potentials for the inflaton field in our analysis. The first is the $\alpha$-attractor inflation potential
\begin{equation}
\label{th}
	V(\phi)=V_0 \operatorname{th}^2\Big(\frac{\phi}{\sqrt{6\alpha}M_{\mathrm{P}}}\Big)
\end{equation}
with $\alpha=1$, $V_0=10^{-10}M_{\mathrm{P}}^4$ and the second is the quadratic potential
\begin{equation}
\label{quadratic}
	V(\phi)=\frac{m_{\phi}^2 \phi^2}{2}
\end{equation}
with $m_{\phi}=6\!\times\! 10^{-6} M_{\mathrm{P}}$. The amplitudes of both potentials were chosen from the requirement that they imply the correct amplitude of the curvature power spectrum constrained by the CMB observations~\cite{Planck:2018-infl}.
\footnote{Here we disregard the fact that the quadratic inflaton potential~\eqref{quadratic} is strongly disfavored by the CMB observations~\cite{Planck:2018-infl} and use it for illustrative purposes. Moreover, in many viable inflationary models, when the inflaton approaches the minimum of its potential, the latter can be well approximated by the $\phi^2$ term.}
The axial coupling function has the simplest linear form for both potentials
\begin{equation}
	I(\phi)=\frac{\beta\phi}{M_{\mathrm{P}}}\, ,
\end{equation}
where $\beta$ is the dimensionless coupling constant typically varying in the range $10-30$.

The initial conditions for the inflaton and its derivative are given by
\begin{equation}
	\phi(0)=6.25\, M_{\mathrm{P}}\, , \qquad \dot{\phi}(0)=-\frac{M_{\mathrm{P}}V'(\phi_0)}{\sqrt{3V(\phi_0)}}\approx -1.13\!\times\! 10^{-7}M_{\mathrm{P}}^2\, ,
\end{equation}
for potential~\eqref{th} and
\begin{equation}
	\phi(0)=15.55\, M_{\mathrm{P}}\,, \qquad \dot{\phi}(0)=-\frac{M_{\mathrm{P}}V'(\phi_0)}{\sqrt{3V(\phi_0)}}\approx -4.9\!\times\! 10^{-6}M_{\mathrm{P}}^2\, ,
\end{equation}
for potential~\eqref{quadratic}, where the inflaton initial value allows us to get at least 60 $e$-foldings of inflation and the value of its initial derivative is computed assuming the slow-roll approximation. Note that only the last 10–15 $e$-foldings are important for magnetogenesis and fermion production; however, the initial conditions should be specified well before this moment. The initial conditions for gauge-field bilinear functions, energy density, chirality density, and the number density of produced particles are set to zero.

\begin{figure}[ht]
	\centering
	\includegraphics[height=4.1 cm]{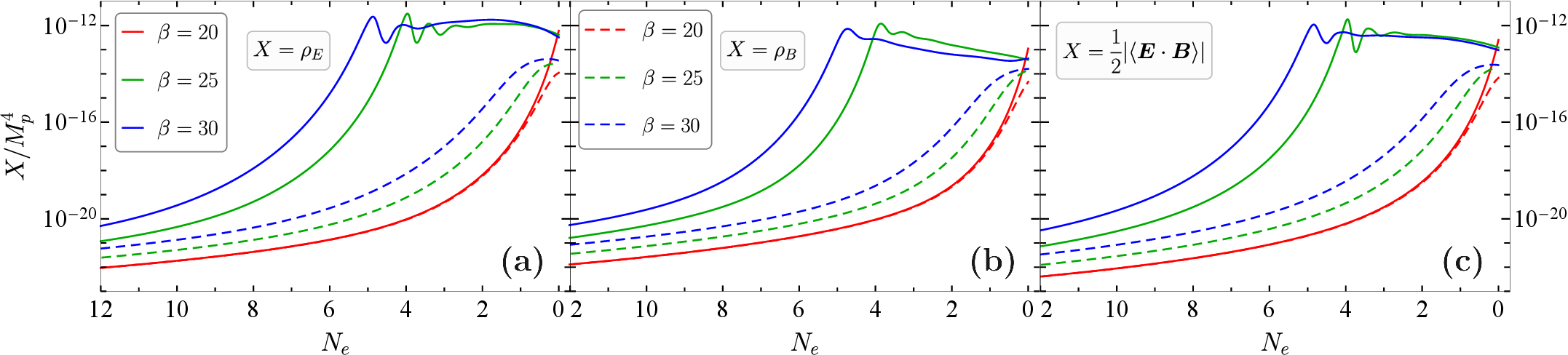}
	\caption{The electric energy  density $\rho_{E}$ [panel~(a)], magnetic energy density $\rho_{B}$ [panel~(b)], and Chern-Pontryagin density $\frac{1}{2}|\langle \boldsymbol{E}\cdot \boldsymbol{B}\rangle|$ [panel~(c)] as functions of the number of $e$-foldings counted from the end of inflation $N_{e}$ for the $\alpha$-attractor potential~\eqref{th} and three different values of the axial coupling parameter $\beta=20$ (red lines), $\beta=25$ (green lines), and $\beta=30$ (blue lines). The dependence for weakly interacting particles ($k=1)$ is shown by dashed lines and by solid lines for strongly interacting particles ($k=10^4$).}
	\label{fig-1}
\end{figure}

\begin{figure}[ht]
	\centering
	\includegraphics[width=7 cm]{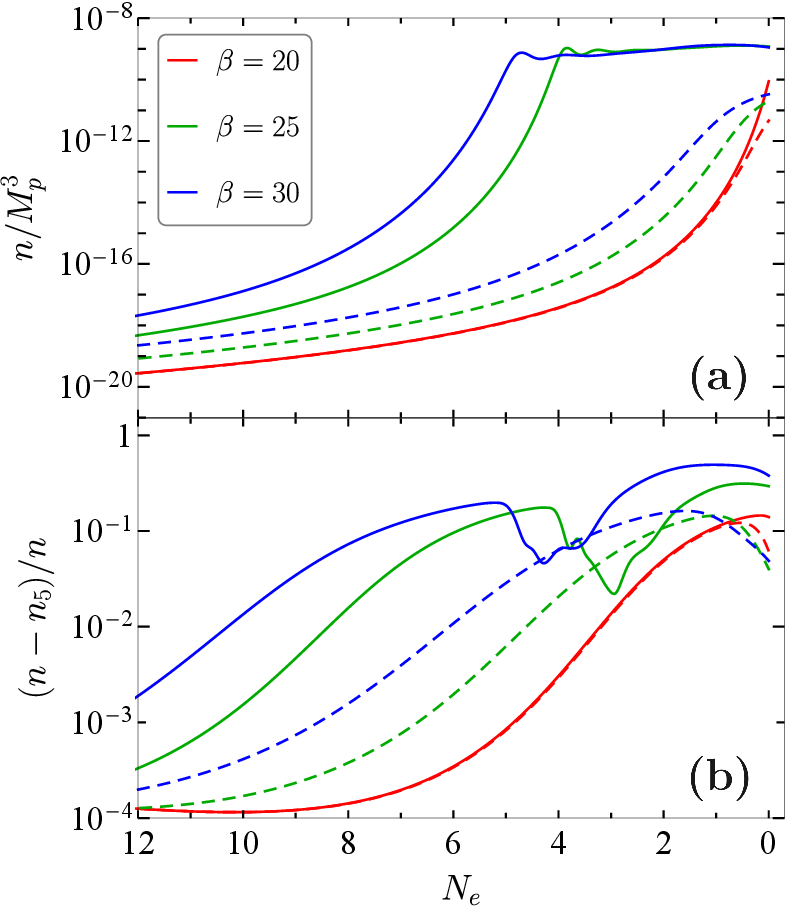}
	\caption{The total number density $n$ [panel~(a)] and chiral asymmetry $(n-n_5)/n$ [panel~(b)] as a function of the number of $e$-foldings counted from the end of inflation $N_{e}$ for the $\alpha$-attractor potential~\eqref{th} and three different values of the axial coupling parameter $\beta=20$ (red lines), $\beta=25$ (green lines), and $\beta=30$ (blue lines). The dependence for weakly interacting particles ($k=1)$ is shown by dashed lines and by solid lines for strongly interacting particles ($k=10^4$).}
	\label{fig-2}
\end{figure}

\begin{figure}[ht]
	\centering
	\includegraphics[width=7 cm]{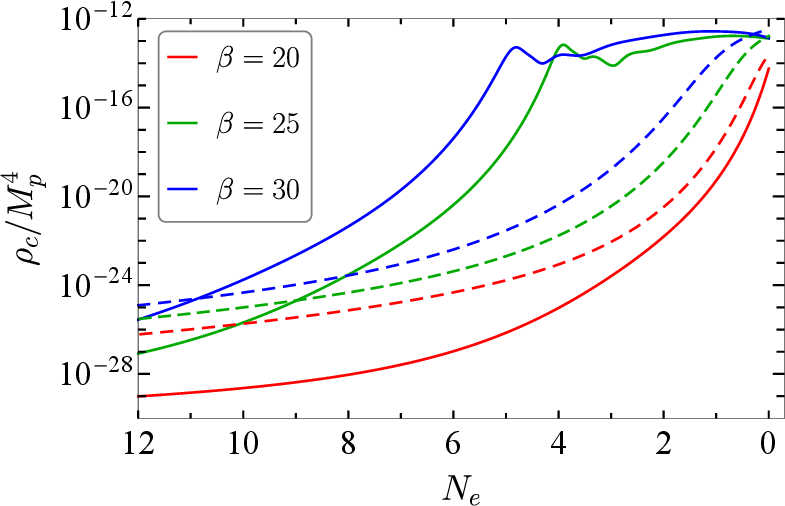}
	\caption{The dependence of the fermion energy density $\rho_{\mathrm{c}}$ on the number of $e$-foldings counted from the end of inflation $N_{e}$ in the case of the $\alpha$-attractor potential~\eqref{th} and	three values of the axial coupling parameter $\beta=20$ (red lines), $\beta=25$ (green lines), $\beta=30$ (blue lines). The dependence for weakly interacting particles ($k=1)$ is shown by dashed lines and by solid lines for strongly interacting particles ($k=10^4$).}
	\label{fig-3}
\end{figure}

Our numerical analysis revealed that the Hubble time is much less than the collision time $\tau_H\ll\tau_{\mathrm{p}}$ for weakly interacting particles during the whole inflation; hence, the corresponding plasma is in the collisionless regime and the expressions for conductivities~\eqref{sigmaE_D} and~\eqref{sigmaB_D} should be applied. For strongly interacting particles, vice versa, the collision time is much less than the Hubble time  $\tau_H\gg\tau_{\mathrm{p}}$ during inflation, \textit{i.e.}, this plasma in the collision regime; hence, the expressions for conductivities~\eqref{sigmaE} and~\eqref{sigmaB} should be used throughout the whole inflation.

\begin{figure}[ht]
	\centering
	\includegraphics[height=4.1 cm]{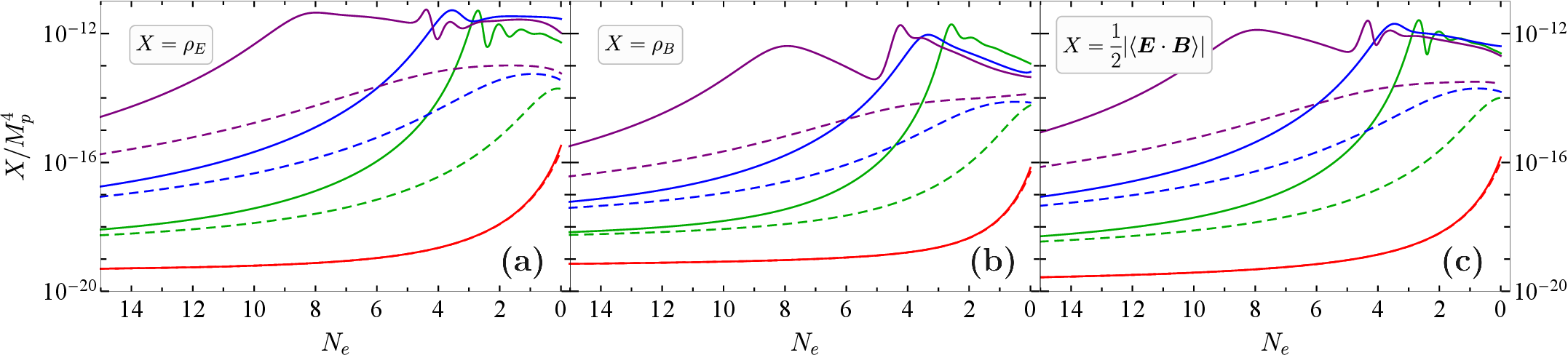}
	\caption{The same dependences as shown in Fig.~\ref{fig-1} for the quadratic potential~\eqref{quadratic} and four values of the axial coupling parameter $\beta=10$ (red lines), $\beta=15$ (green lines), $\beta=20$ (blue lines), $\beta=25$ (purple lines).}
	\label{fig-4}
\end{figure}
\begin{figure}[ht]
	\centering
	\includegraphics[width=7 cm]{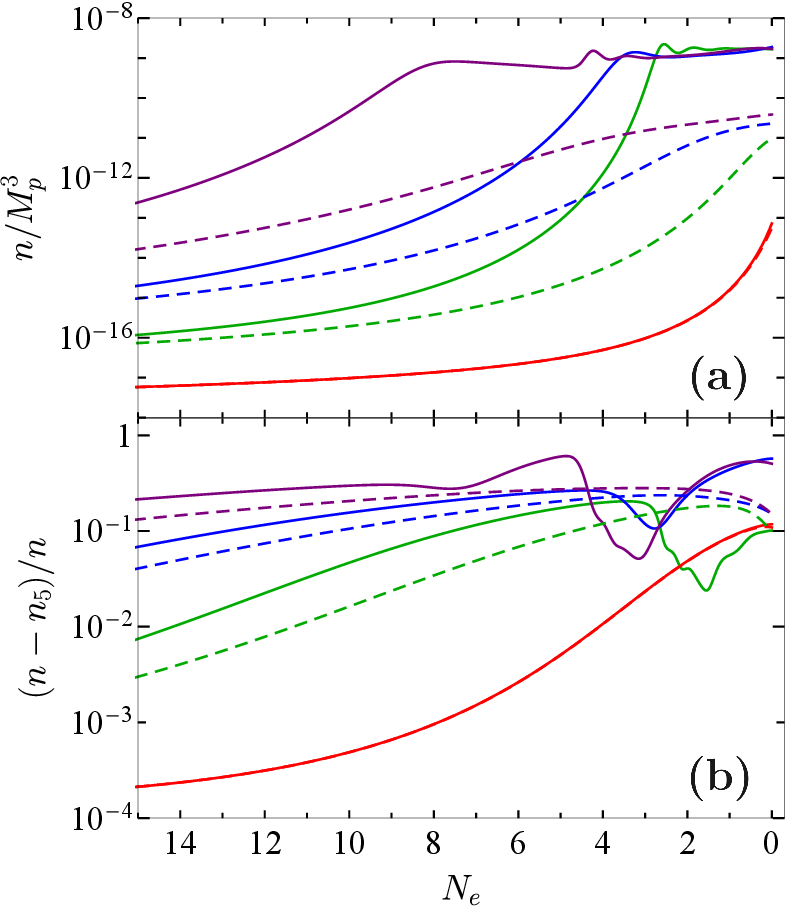}
	\caption{The same dependences as shown in Fig.~\ref{fig-2} for the quadratic potential \eqref{quadratic} and four different values of the axial coupling parameter $\beta=10$ (red lines), $\beta=15$ (green lines), $\beta=20$ (blue lines), $\beta=25$ (purple lines).}
	\label{fig-5}
\end{figure}
\begin{figure}[ht]
	\centering
	\includegraphics[width=7 cm]{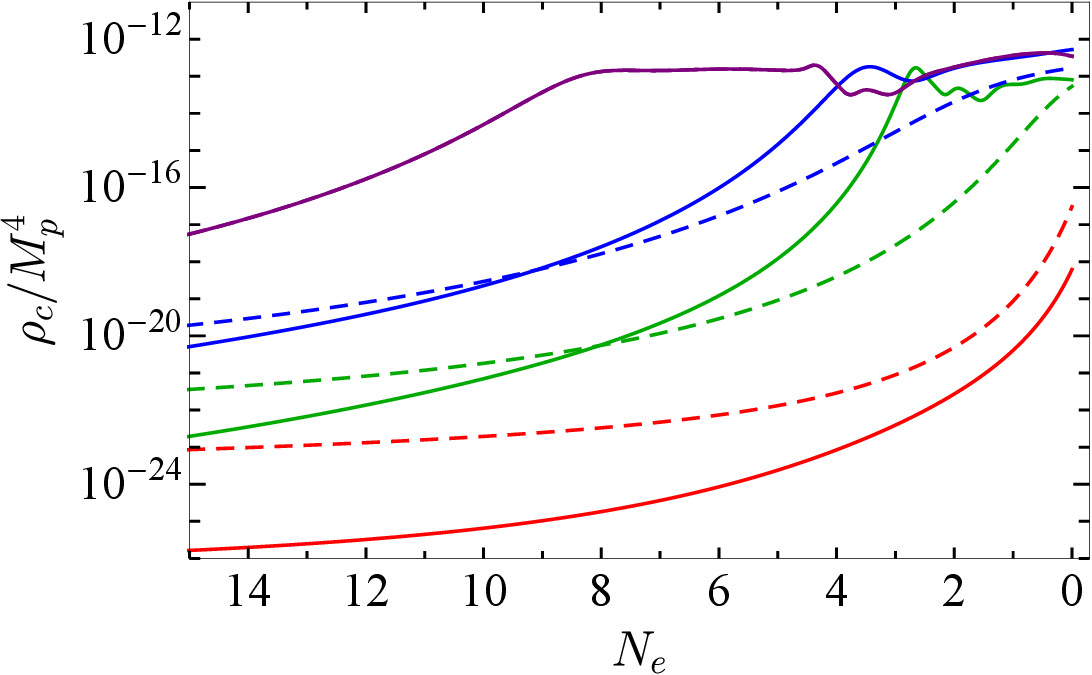}
	\caption{The same dependences as shown in Fig.~\ref{fig-3} for the quadratic potential \eqref{quadratic} and four different values of the axial coupling parameter $\beta=10$ (red lines), $\beta=15$ (green lines), $\beta=20$ (blue lines), $\beta=25$ (purple lines).}
	\label{fig-6}
\end{figure}

The most general results of our analysis could be summarized as follows. According to Figs.~\ref{fig-1}, \ref{fig-2}(a), \ref{fig-4}, \ref{fig-5}(a), as one could expect, the generated electric, magnetic, Chern-Pontryagin, and fermion number densities at a given value of $\beta$ are larger for strongly interacting particles compared to the case of weakly interacting particles because smaller collision time reduces the conduction electric current which tends to screen the electric field. This makes generated electromagnetic fields stronger. On the other hand, no such a universal conclusion could be drawn for the generated fermion energy density as Figs.~\ref{fig-3} and \ref{fig-6} imply. As to chiral asymmetry $(n-n_5)/n$, its value at the end of inflation at a given value of $\beta$ like the value of generated electromagnetic fields is larger for strongly interacting particles compared to the case of weakly interacting particles.

As to the role of the axial coupling constant $\beta$, Figs.~\ref{fig-1}--\ref{fig-6} demonstrate a nonmonotonic dependence of the generated electromagnetic fields, fermion number and energy densities, and chiral asymmetry on the number of $e$-foldings $N_e$ from the end of inflation for a sufficiently large value of the axial coupling constant $\beta$.

Comparing Figs.~\ref{fig-1}, \ref{fig-2}, \ref{fig-3} and Figs.~\ref{fig-4}, \ref{fig-5}, \ref{fig-6} plotted for the $\alpha$-attractor and quadratic potentials, respectively, we see that the corresponding results are rather similar, \textit{i.e.}, the characteristics of generated electromagnetic fields and chiral asymmetry do not show a significant dependence on the form of the inflaton potential.

Let us provide more detailed quantitative information on the obtained results. For weakly interacting particles in the case of the $\alpha$-attractor potential~\eqref{th},  the increase of coupling constant $\beta$ from $20$ to $30$ results in the increase of the electric and magnetic field energy densities and Chern-Pontryagin density $\frac{1}{2}|\langle \boldsymbol{E}\cdot \boldsymbol{B}\rangle|$ (see, Fig.~\ref{fig-1}) as well as in the increase of the fermion number (see, Fig.~\ref{fig-2}) and energy densities (see, Fig.~\ref{fig-3}). The chiral asymmetry $(n-n_5)/n$ is equal to $0.062$, $0.040$ and $0.049$ for $\beta=20$, $25$ and $30$, respectively, i.e., it is not monotonic as $\beta$ changes. 

For strongly interacting particles, where the collisional regime is realized due to strong interactions, the magnitude of the considered densities is more than ten times larger than that in the case of weakly interacting particles. In addition, chiral asymmetry increases to $0.14$, $0.29$, and $0.38$ and is monotonic with $\beta$. Moreover, qualitative changes in the time evolution are observed for $\beta=25$ and $30$, namely, a non-monotonic behavior due to the backreaction of produced gauge fields. Indeed, while the highest value of the magnetic energy density is observed for $\beta=15$, the other considered quantities reach their highest values for $\beta=20$. The largest value of chiral asymmetry is $0.57$.

For the quadratic potential~\eqref{quadratic}, the increase of $\beta$ from $10$ to $25$ also leads to the increase of $\rho_E$, $\rho_B$ as well as  $\frac{1}{2}|\langle \boldsymbol{E}\cdot \boldsymbol{B}\rangle|$, $n$, and $\rho_{\mathrm{c}}$ (see, Figs.~\ref{fig-4}--\ref{fig-6}). The most considerable growth is observed between $\beta=10$ and $15$. The value of $(n-n_5)/n$ is in the range from $0.095$ to $0.15$.

\section{Summary}
\label{sec:summary}

Our analysis of chirality production and its impact on generated electromagnetic fields during axion inflation via the gradient-expansion formalism and hydrodynamical approach (taking into account particle collisions in the relaxation time approximation) in the gauge field and fermion sectors, respectively, led to the following results.

Comparing the particle collision time with the Hubble time, we found that local thermodynamic equilibrium is not reached for weakly interacting particles equilibrating only via the electroweak interactions and a realistically small number of particle species. However, for strongly interacting particles, the characteristic collision time appears to be much smaller than the Hubble time. Although the intense particle production due to the Schwinger effect may still prevent the system from reaching the state of local thermodynamic equilibrium, the particle collisions may have a strong impact on the pair production process and, consequently, on the outcome of magnetogenesis during pseudoscalar inflation.

We found that the generated electric, magnetic, Chern-Pontryagin, and fermion number densities at a given value of the coupling constant $\beta$ are larger for strongly interacting particles compared to the case of weakly interacting particles as one could expect because smaller collision time reduces the conduction electric current which tends to decrease the electric field. Therefore, generated electromagnetic fields are stronger. Although the value of produced chiral asymmetry at the end of inflation at fixed $\beta$ is larger for strongly interacting particles compared to the case of weakly interacting particles, chiral asymmetry can be larger for weakly interacting particles for a few $e$-foldings close to the end of inflation. For a sufficiently large value of the axial coupling constant $\beta$, a nonmonotonic dependence of generated electromagnetic fields, fermion number and energy densities, and chiral asymmetry on the number of $e$-foldings is observed near the end of inflation due to the strong backreaction of produced gauge fields. In addition, the obtained results show that the values of generated electromagnetic fields and chiral asymmetry do not depend notably on the form of the inflaton potential.

We would like to note that the present study allows us to draw only some general qualitative conclusions about chirality production during axion inflation and does not claim to provide an accurate quantitative description of the process. The latter could be realized only in the framework of the chiral kinetic theory with a realistic collision integral describing the interaction processes in plasma. We plan to address this issue elsewhere.

\begin{acknowledgments}	
	The authors are grateful to  S.I.~Vilchinskii for useful discussions and participation in the early stage of this project. The work of E.V.G., A.I.M., and O.M.T. was supported by the National Research Foundation of Ukraine Project No.~2020.02/0062. 	
\end{acknowledgments}

\end{document}